\documentclass{article}
\usepackage{spconf,amsmath,graphicx}
\usepackage{cite}
\usepackage{subcaption}
\usepackage{multirow}
\title{Multi-resolution fully convolutional neural networks for Monaural audio source separation}
\name{Emad M. Grais, Hagen Wierstorf, Dominic Ward, and Mark D. Plumbley}% %\thanks{Emad M. Grais is also with Helwan university, Egypt.}}
\address{Centre for Vision, Speech and Signal Processing \\ University of Surrey, Guildford, UK.}
\begin{document}
\ninept
\maketitle
\begin{abstract}
In deep neural networks with convolutional layers, each layer typically has fixed-size/single-resolution receptive field (RF). Convolutional layers with a large RF capture global information from the input features, while layers with small RF size capture local details with high resolution from the input features. In this work, we introduce novel deep multi-resolution fully convolutional neural networks (MR-FCNN), where each layer has different RF sizes to extract multi-resolution features that capture the global and local details information from its input features. The proposed MR-FCNN is applied to separate a target audio source from a mixture of many audio sources. Experimental results show that using MR-FCNN improves the performance compared to feedforward deep neural networks (DNNs) and single resolution deep fully convolutional neural networks (FCNNs) on the audio source separation problem.   
\end{abstract}
%The objects and patterns of different audio sources in a spectrogram of a mixture of many  audio signals usually have different spectro-temporal sizes. 
\section{Introduction}
\label{sec:intro}
Monaural audio source separation (MASS) aims to separate audio sources from their mono/single mixture \cite{xiao:16:delmss,grais:14:ssrnmfmmse,Virtanen:07:msssbnmfwtcasc}. Many deep learning techniques have been used before to tackle this problem \cite{Mimilakis:17:redasfcmsvs,grais:17:tsscass,Yuxuan:14:aspttsss,Grais:17:descassdnn}. %For their ability in extracting robust spectro-temporal structures/patterns of different audio signals \cite{Honglak:09:uflaccdbn,Jonathan:11:scaehfe,Bo:17:scdaefr}, convolutional neural networks (CNN) have been used successfully to learn useful features in many audio processing applications \cite{Yanmin:16:vdcnnnrsr,Szu:16:snrcnnmse,Yong:17:cgrnnisfat,Yoonchang:17:dcnnpirpm,Keunwoo:16:crnnmc,Filip:16:fcdammcr}. 

A variety of deep neural networks with convolutional layers have been used recently to tackle the MASS problem \cite{chandna:17:massudcnn,Venkataramani:17:nnacamss,Venkataramani:17:eessafe,grais:17:scasscda,Miron:17:msisscmcnn,Lim:17:hpsscae,fu:17:eewuedemofcnn}. One of the main differences in those works relies on using either fully convolutional neural networks (FCNN), where all the network layers are convolutional layers, or some of the layers are convolutional and others are fully connected layers. The common aspect in those works is that each convolutional layer composes of a set of filters that have the same receptive field (RF) size. The RF is the field of view of a unit (filter in the FCNN case) in a certain layer in the network \cite{Wenjie:16:uerfdcnn}. In the fully connected neural networks (DNN), the output of each unit in a certain layer depends on the entire input to that layer, while the output of a unit in a convolutional layer only depends on a region of the input, this region in the input is the RF for that unit. The RF size is a crucial issue in many audio and visual tasks, as the output must respond to areas with sizes correspond to the sizes of the different objects/patterns in the input data to extract useful information/features about each object \cite{Wenjie:16:uerfdcnn}. 
The size of the RF equals the size of the filters in a convolutional layer. A large filter size captures the global structure of its input features \cite{Kawahara:16:mrtchpsltl,tang:12:mdbn}. A small filter size captures the local details with high resolution but it does not capture the global structure of its input features. Intuitively, it might be useful to have sets of filters that can extract both the global structures and local details from the input features in each layer. This might be useful in MASS problem, since the input signal is a mixture of different audio sources and useful features can be extracted for certain sources in certain time-frequency resolutions which may differ from one source to another. %in other words, the objects and patterns of different audio sources in a spectrogram of a mixture of many  audio signals usually have different spectro-temporal sizes.   

The concept of extracting multi-resolution features has been proposed recently in many applications with different ways of extracting and combining the multi-resolution features from the input data \cite{Zhang:17:HeartID,Xue:17:emrtscar,Naderi:17:mrcnnrsr,Kawahara:16:mrtchpsltl}. In this paper, we introduce a novel multi-resolution fully convolutional neural network (MR-FCNN) model for MASS, where each layer in the MR-FCNN is a convolutional layer that is composed of different sets of filters with different sizes to extract the global and local information from its input features in each layer in different resolutions. 
Each set of filters has filters with the same size which is different than the sizes of the filters in the other sets. We believe that, this is the first time that a deep neural network has been proposed with each layer composed of multi-resolution filters that extract multi-resoltuion features from the layer before, and it is the first time the concept of extracting multi-resolution features is used for MASS problem. The inputs and outputs of the MR-FCNN are two-dimensional (2D) segments from the magnitude spectrogram of the mixed and target source signals respectively. The MR-FCNN in this work is trained to extract useful spectro-temporal features and patterns in different time-frequency resolutions to separate the target source from the input mixture.   

This paper is organized as follows: Section \ref{fcnn-mr-fcnn} shows a brief introduction about the fully convolutional neural networks and the proposed MR-FCNN. The proposed approach of using MR-FCNN for MASS is presented in Section \ref{overall}. The rest of the paper is for the experiments, discussions, and conclusions.

\section{Multi-resolution fully convolutional neural networks}
\label{fcnn-mr-fcnn}
In this section we first give an introduction about the fully convolutional neural network (FCNN) that we use in this study as a core model and then we introduce the proposed MR-FCNN.
\subsection{Fully convolutional neural networks}
\label{fcnn}
\begin{figure}[h]
 \includegraphics[width=1.05\linewidth,height=3.5cm]{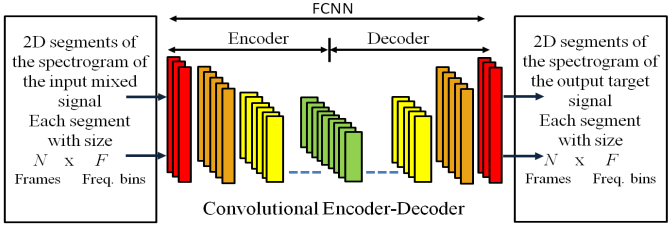}%chart1_email
\caption{\label{fig:fcnn}{\scriptsize{The overview of the structure of a FCNN that separates one target source from the mixed signal. Each layer consists of a single set of filters with the same size followed by a rectified linear unit (ReLU) as activation function. The set of filters in the input and output layers have large filter sizes and small number of filters. The number of filters increases and the size decreases when getting further from the input and output layers \cite{se:16:fcnnse}. There is symmetric in the filter sizes and numbers of filters between the encoder and decoder sides.}}}
%\end{center}
\end{figure}

The FCNN model that is used here is somewhat similar to the convolutional denoising encoder-decoder (auto-encoder) networks (CDEDs) that was used in \cite{grais:17:scasscda,se:16:fcnnse}, but without using either down-sampling (pooling) or up-sampling as shown in Fig. \ref{fig:fcnn}.   %Pooling \cite{Dominik:10:epocaor} was shown to not work well on a similar regression problems to MASS \cite{se:16:fcnnse,Szu:16:snrcnnmse}. 
The encoder part in the FCNN is composed of repetitions of a convolutional layer and an activation layer. Each convolutional layer consists of a set of filters with the same size to extract features from its input layer, the activation layer is the rectified linear unit (ReLU) that imposes nonlinearity to the feature maps. The FCNN is trained from corrupted input signals and the encoder part is used to extract noise robust features that the decoder can use to reconstruct a cleaned-up version of the input data \cite{se:16:fcnnse,Mengyuan:16:mrcdasr}. In MASS, the input mixed signal can be seen as a sum of the target source that needs to be separated and background noise (the other sources in the mixture). The decoder part consists of repetitions of a deconvolutional (transposed convolution) layer and an activation layer. The input and output data are 2D signals (magnitude spectrograms) and the filtering is a 2D operator.

\subsection{MR-FCNN}
\label{mr-fcnn}
Each layer in the FCNN in Fig.\ref{fig:fcnn} is composed of one set of filters that have the same RF size. The size of the RF is a very important parameter as the output of each filter must respond to areas with sizes correspond to the sizes of the different objects/patterns in the input to extract useful information/features from the input data \cite{Wenjie:16:uerfdcnn}. For example, if the size of the RF of a filter is much bigger than the size of the input pattern, the filter will capture blurred features from the input patterns, while if the RF of a unit is smaller than the size of the input pattern, the output of the filter loses the global structure of the input pattern \cite{Wenjie:16:uerfdcnn}. 

\begin{figure}[h]
 \includegraphics[width=1.1\linewidth,height=6.5cm]{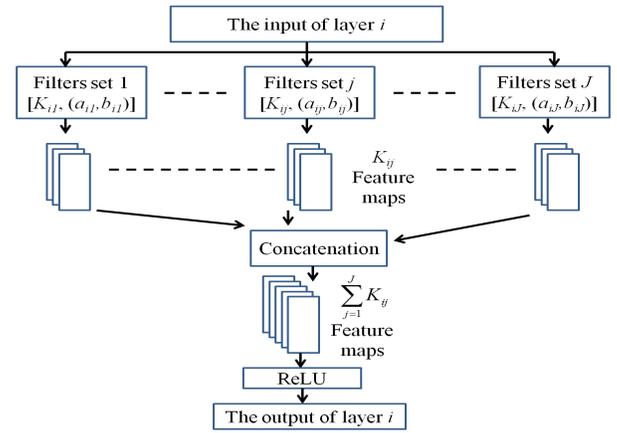}%chart1_email
\caption{\label{fig:layer_mrfcnn}{\scriptsize{The overview of the proposed structure of each layer of the MR-FCNN. $K_{ij}$ denotes the number of filters with size $a_{ij} \times b_{ij}$ in set $j$ in layer $i$. $a_{ij}$ is the dimension in the time direction of the filters and $b_{ij}$ is the dimension in the frequency direction of the filters in set $j$ and layer $i$. The filters in different sets have different sizes and the filters within a set have the same size. Each set $j$ in layer $i$ generates $K_{ij}$ feature maps. The number of feature maps that each layer $i$ generates equal to the sum of the number of feature maps that all the sets in layer $i$ generate ($\sum_{j=1}^J K_{ij}$). ReLU denotes a rectified linear unit (ReLU) as an activation function. 
}}}
\end{figure}

In audio source separation problems, the spectrogram of the input mixed signal usually contains different combinations of different spectro-temporal patterns from different audio signals. There are unique set of patterns associated with each source in the spectrogram of their mixture and these patterns appear in different spectro-temporal sizes and these sizes are source dependent \cite{Klapuri:07:spmmt}. So, to use the FCNN to extract useful information about the individual sources in the spectrogram of their mixture, it might be useful to use filters with different RF sizes in each layer, where the different RF sizes are proportional to the diversity of the spectro-temporal sizes of the patterns in the spectrogram. 
%
%If the filter has large size, its RF covers a large area from its input features. This is good in extracting global information about the input data. If the filter size is small, it captures local information in high resolution but it can not capture the correlation between the elements of the input data in large region. 
%
%In the case of having the input data as a magnitude spectrogram of audio signals, it may be useful to obtain global information about the correlation between the spectrogram elements in large time-frequency segments, but it is also useful to capture the local details in high resolution of the audio patterns that appear in the spectrogram. Also, you can obtain better features from certain audio signals using operators with certain time-frequency resolution than using other resolutions. In MASS problem, the input signal is a mixture of different audio sources and useful features can be extracted for certain sources in certain time-frequency resolutions that is different than the suitable time-frequency resolutions for the other sources. 
Bearing these issues in mind, we propose MR-FCNN which is the FCNN shown in Fig.\ref{fig:fcnn} but with multi-resolution filters (filters with different sizes) in each layer. Thus, each layer in the MR-FCNN has sets of 2D filters. Each set of filters has the same size which is different than the size of the filters in the other sets in the same layer. Each set of filters generates feature maps with certain time-frequency resolution. Fig. \ref{fig:layer_mrfcnn} shows the detail structure for each layer in the MR-FCNN. Each layer in the MR-FCNN generates multi-resolution features from its input features and also combines the multi-resolution features from the previous layers to generate accurate patterns that compose the structure of the underlying data. %Note that, in FCNN each layer has filters with the same size, while in MR-FCNN each layer has sets of filters where each set has filters with the same size that is different than the size of the filters in the other sets.  
% It maybe useful to obtain the global information the gives an idea about how the elements of the input data in large area interact with each other. Global details help in capturing the harmonic structure in the frequency direction and the smoothness in the time direction of the spectrogram. 
\section{MR-FCNN for MASS}
\label{overall}
Given a mixture of $L$ sources as $y(t) = \sum_{l=1}^L s_l(t)$, the aim of MASS is to estimate the sources $s_l(t), \ \forall{l}$, from the mixed signal $y(t)$ \cite{emad:12:avsrwbmscss,emad:13:stpemenbscss}. We work here in the short-time Fourier transform (STFT) domain. Given the STFT of the mixed signal $y(t)$, the main goal is to estimate the STFT of each source in the mixture.

In this work, we propose to use as many MR-FCNN as the number of sources to be separated from the mixed signal. Each MR-FCNN sees the mixed signal as a combination of its target source and background noise. The main aim of each MR-FCNN is to estimate a clean signal for its corresponding source from the other background sources that exist in the mixed signal. This is a challenging task for each MR-FCNN since each MR-FCNN deals with highly nonstationary background noise (other sources in the mixture). Each MR-FCNN is trained to map the magnitude spectrogram of the mixture into the magnitude spectrogram of its corresponding target source. Each MR-FCNN in this work is a deep fully 2D multi-resolutional convolutional neural network without any fully connected layer, which keeps the number of parameters to be optimized for each MR-FCNN small. Also using fully 2D convolutional layers allows neat 2D spectro-temporal representations for the data through all the layers in the network. The inputs and outputs of the MR-FCNNs are 2D-segments from the magnitude spectrograms of the mixed and target signals respectively. Therefore, the MR-FCNNs span multiple spectral frames to capture the spectro-temporal characteristics of each source. The number of spectral frames that each input segment has is $N$ and the number of frequency bins is $F$. In this work, $F$ is the dimension of the whole spectral frame. 

\subsection{Training the MR-FCNNs for source separation}
\label{sec:train}
Let's assume we have training data for the mixed signals and their corresponding clean/target sources. Let $\mathbf{Y}_{\mbox{tr}}$ be the magnitude spectrogram of the mixed signal and $\mathbf{S}_l$ be the magnitude spectrogram of the clean source $l$. The subscript ``tr'' denotes the training data. The MR-FCNN that separates source $l$ from the mixture is trained to minimize the following cost function:
\begin{equation}
\label{cost_mask}
C_l =\sum_{n,f}\left( \mathbf{Z}_l\left(n,f\right) - \mathbf{S}_l\left(n,f\right) \right)^2
\end{equation}
where $\mathbf{Z}_l$ is the actual output of the last layer of the MR-FCNN of source $l$, $\mathbf{S}_l$ is the reference clean output signal for source $l$, $n$, and $f$ are the time and frequency indices respectively. The input of the MR-FCNNs is the magnitude spectrogram $\mathbf{Y}_{\mbox{tr}}$ of the mixed signal. The input and output instants of the MR-FCNN are 2D-segments, where each segment is composed of $N$ consecutive spectral frames taken from the magnitude spectrograms. This allows the MR-FCNN to learn multi-resolution spectro-temporal patterns for each source. 
\subsection{Testing the MR-FCNNs for source separation}
\label{sec:test} 
Given the trained MR-FCNNs, the magnitude spectrogram $\mathbf{Y}$ of the mixed signal is passed through the trained MR-FCNNs. The output of the MR-FCNN of source $l$ is the estimate $\mathbf{\tilde{S}}_l$ of the spectrogram of source $l$.
\section{Experiments}
\label{sec:exp}
We applied our proposed MASS using MR-FCC approach to separate the voice/vocal sources from a group of songs from the SiSEC-2015-MUS-task dataset \cite{ono:15:tsisec}. The dataset has 100 stereo songs with different genres and instrumentations. To use the data for the proposed MASS approach, we converted the stereo songs into mono by computing the average of the two channels for all songs and sources in the data set. Each song is a mixture of vocals, bass, drums, and other musical instruments. We used our proposed algorithm to separate the vocal signals from each song. 

The first 50 songs in the dataset were used as training and validation datasets to train the MR-FCNN for separation, and the last 50 songs were used for testing. The data were sampled at 44.1kHz. The magnitude spectrograms for the data were calculated using the STFT, a Hanning window with 2048 points length and overlap interval of 512 was used and the FFT was taken at 2048 points, the first 1025 FFT points only were used as features since the conjugate of the remaining 1024 points are involved in the first points.

For the input and output data for the MR-FCNN, we chose the number of spectral frames in each 2D-segment to be 15 frames. This means the dimension of each input and output instant for the MR-FCNN is 15 (time frames) $\times$ 1025 (frequency bins) as in \cite{grais:17:scasscda}. Thus, each input and output instant (the 2D-segments from the spectrograms) spans around 370 msec of the waveforms of the data. 

The quality of the separated sources was measured using the signal to distortion ratio (SDR), signal to interference ratio (SIR), and signal to artefact ratio (SAR) \cite{vincent:06:pmi}. SIR indicates how well the sources are separated based on the remaining interference between the sources after separation. SAR indicates the artefacts caused by the separation algorithm in the estimated separated sources. SDR measures the overall distortion (interference and artefacts) of  the separated sources. The SDR values are usually considered as the overall performance evaluation for any source separation approach \cite{vincent:06:pmi}. Achieving high SDR, SIR, and SAR indicates good separation performance.

We compared the performance of the proposed MR-FCNN model, feedforward deep neural networks (DNNs), and the single resolution FCNN in separating the vocal signals from each song in the test set. The size of each input and output instant is the same in FCNN and MR-FCNN (15$\times$1025). Each input and output instant of the DNN is a single frame of the magnitude spectrograms of the input and output signals respectively. Table \ref{table:cdae} shows the number of layers, the number of filters in each layer, and the size of the filters for the FCNN and MR-FCNN. %The implementation of this work was done using Keras with tensorflow backend \cite{chollet2015keras}

\begin{table}
\scalebox{0.75}
{
\begin{tabular}{||p{1.8cm}|p{2.8cm}|p{0.58cm} p{2.8cm}||}
 \hline
 \multicolumn{4}{|c|}{FCNN and MR-FCNN model summary} \\
 \hline
 \multicolumn{4}{|c|}{The input/output data with size 15 frames and 1025 frequency bins} \\
 \hline
Layer number & FCNN &\multicolumn{2}{|c|}{MR-FCNN} \\
  \hline
\multirow{3}{*}{1} & \multirow{3}{*}{Conv2D[13,(13,21)]} & set 1 & Conv2D[12,(13,21)] \\
                   &                                     & set 2 & Conv2D[3,(7,9)] \\ 
                   &                                     & set 3 & Conv2D[3,(3,3)] \\ 
\hline 
\multirow{3}{*}{2} & \multirow{3}{*}{Conv2D[18,(9,13)]} & set 1 & Conv2D[3,(13,21)] \\
                   &                                     & set 2 & Conv2D[16,(7,9)] \\ 
                   &                                     & set 3 & Conv2D[3,(3,3)] \\ 
                  
\hline
\multirow{3}{*}{3} & \multirow{3}{*}{Conv2D[24,(7,9)]} & set 1 & Conv2D[3,(13,21)] \\
                   &                                    & set 2 & Conv2D[12,(7,9)] \\ 
                   &                                    & set 3 & Conv2D[7,(3,3)] \\ 
  \hline 
\multirow{3}{*}{4} & \multirow{3}{*}{Conv2D[42,(3,3)]} & set 1 & Conv2D[3,(13,21)] \\
                   &                                    & set 2 & Conv2D[3,(7,9)] \\ 
                   &                                    & set 3 & Conv2D[32,(3,3)] \\ 
  \hline 
\multirow{3}{*}{5} & \multirow{3}{*}{Conv2D[24,(7,9)]} & set 1 & Conv2D[3,(13,21)] \\
                   &                                    & set 2 & Conv2D[12,(7,9)] \\ 
                   &                                    & set 3 & Conv2D[7,(3,3)] \\   
  \hline 
\multirow{3}{*}{6} & \multirow{3}{*}{Conv2D[18,(9,13)]} & set 1 & Conv2D[3,(13,21)] \\
                   &                                    & set 2 & Conv2D[16,(7,9)] \\ 
                   &                                    & set 3 & Conv2D[3,(3,3)] \\
  \hline 
\multirow{3}{*}{7} &\multirow{3}{*}{Conv2D[13,(13,21)]}& set 1 & Conv2D[12,(13,21)] \\
                   &                                    & set 2 & Conv2D[3,(7,9)] \\ 
                   &                                    & set 3 & Conv2D[3,(3,3)] \\
  \hline
8                  &   Conv2D[1,(15,1025)]       & \multicolumn{2}{|c||}{Conv2D[1,(15,1025)]} \\
  \hline
total number of parameters & 445,173 & \multicolumn{2}{|c||}{558,181} \\
  \hline
% \multicolumn{3}{|c|}{The output data with size 15 frames and 1025 frequency bins} \\
 \hline
\end{tabular}
}
\caption{\scriptsize{The detail information about the number and sizes of the filters in each layer. For example ``Conv2D[13,(13,21)]'' denotes 2D convolutional layer with 13 filters and the size of each filter is 13$\times$21 where 13 is the size of the filter in the time-frame direction and 21 in the frequency direction of the spectrogram.}}
\label{table:cdae} % is used to refer this table in the text
\end{table}

As in many deep learning models, there are many parameters in the proposed MR-FCNN to be chosen (number of layers, filter size, and the number of filters in each set) and usually these choices are data and application dependent. Choosing the parameters for the FCNN is also not easy. In this work, we follow the same strategy as in \cite{se:16:fcnnse} where the size of the filter is decreasing and the number of filter is increasing when we go deep in the encoder part and the opposite (the filter size increases and the number of the filter decreases) in the decoder part in the output direction. For MR-FCNN, the number and size of the filters in each set in each layer are need to be decided. We restricted ourself in this work to use only three sets of filters for the whole network. The first set with size 13$\times$21, the second set with size 7$\times$9, and the third set with size 3$\times$3. Which means each layer has sets of filters with three different resolutions. Also following the same concept in \cite{se:16:fcnnse} for choosing the number of filters, the layers towards the input and output layers have more filters with large size than the layers in the middle. The layers in the middle have more filters in the set with small filter size than the layers toward the input and output layers. For example, the first layer in MR-FCNN has a set of 12 filters with size 13$\times$21, a set of 3 filters with size 7$\times$9, and a set of 3 filters with size 3$\times$3. Thus, the first layer generates 18 feature maps with three different resolutions. Each feature mape is 15$\times$1025 (the same size of the input and output data). The DNN has three hidden layers with ReLU as activation functions. Each hidden layer has 1025 nodes. The parameters of the DNN are tuned based on our previous work on the same dataset \cite{Emad:16:scassdnne, grais:16:cmescassdnn}. The DNN here has 4,206,600 parameters, the FCNN has 445,173 parameters, and the MR-FCNN has 558,181 parameters.

The parameters for all the networks were initialized randomly. All the networks were trained using backpropagation with gradient descent optimization
using Adam \cite{adam:14:amso} with parameters: $\beta_1=0.9$, $\beta_2=0.999$, $\epsilon=1e-08$, batch size 100, and a learning rate starts with $0.0001$ and reduced by a factor of 10 when the values of the cost function do not decrease on the validation set for 3 consecutive epochs. The maximum number of epochs is 100. We implemented our proposed algorithm using Keras with Tensorflow backend \cite{chollet2015keras}.

\begin{figure}

\centering
   \begin{subfigure}[b]{0.55\textwidth}
   \includegraphics[width=0.9\linewidth, height=6.5cm]{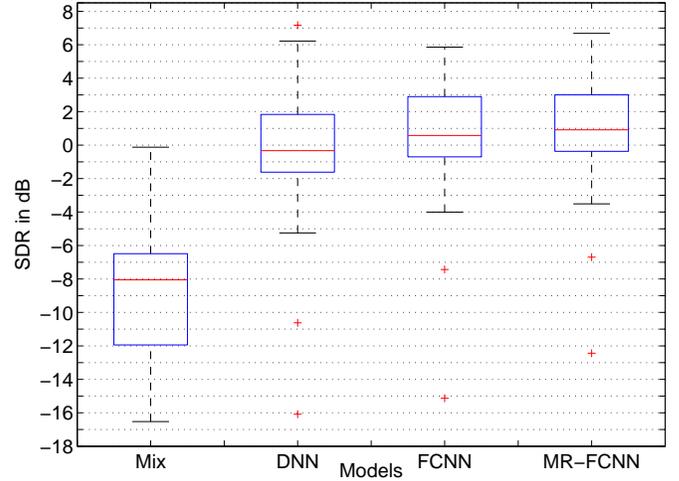} 
   \caption{SDR in dB}\hfill \hfill\hfill\hfill\hfill\hfill
   \label{fig:dnn1} 
\end{subfigure}

\begin{subfigure}[b]{0.55\textwidth}
   \includegraphics[width=0.9\linewidth, height=6.5cm]{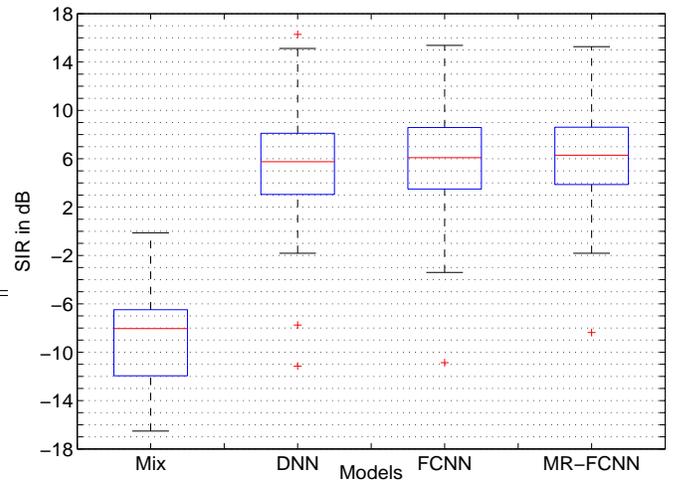} 
   \caption{SIR in dB}\hfill\hfill\hfill\hfill\hfill\hfill 
   \label{fig:dnn2}
\end{subfigure}

\begin{subfigure}[b]{0.55\textwidth}
   \includegraphics[width=0.9\linewidth, height=6.5cm]{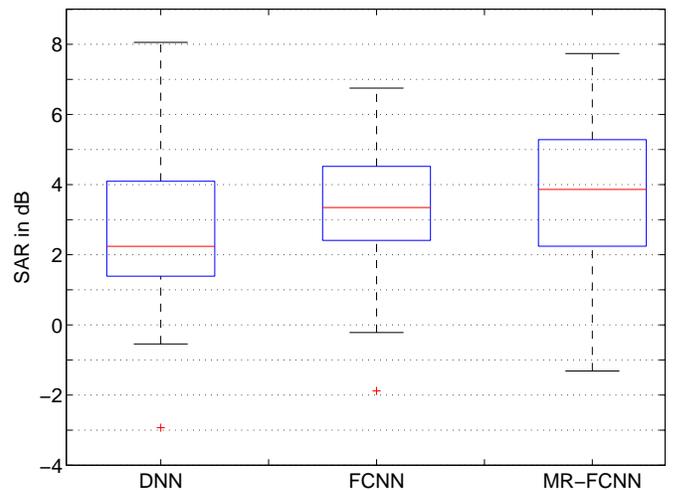}
   \caption{SAR in dB}\hfill\hfill
   \label{fig:dnn3} 
\end{subfigure}

\caption[]{\scriptsize{(a) The SDR, (b) the SIR, and (c) the SAR (values in dB) for the separated vocal signals of using deep fully connected feedforward neural networks (DNNs), using deep fully convolutional neural networks (FCNNs), and the proposed multi-resolution fully convolutional neural networks (MR-FCNN). ''Mix`` denotes the input mixed signal.}}
\label{fig:dnn_dnn} 
\end{figure}

To compare the proposed MR-FCNN model to the FCNN, we tried to adjust the number of filters and their sizes in each layer of both models to have total number of parameters in both models close to each other as shown in Table \ref{table:cdae}. Fig.\ref{fig:dnn_dnn}, shows the box-plot of the SDR, SIR, and SAR of the separated vocal sources using three different deep learning models, namely DNN, FCNN, and MR-FCNN. The figure also shows the SDR and SIR values of the target vocal source in the mixed signal (denoted as Mix in Fig.\ref{fig:dnn_dnn}). We did not show the SAR of the mixed signal because it is usually very high (around 250 dB) and causes scaling problem in the figure. From the figure we can see that the vocal signals in the input mixed signal (denoted as Mix in Fig.\ref{fig:dnn_dnn}) have very low SDR and SIR values, which shows that we are dealing with a very challenging source separation problem.

As can be seen from Fig.\ref{fig:dnn_dnn}, the three methods perform well on the SDR, SIR, and SAR values of the separated vocal signals. The proposed MR-FCNN model outperforms the two other models in the SDR and SAR values. All the models perform similarly in the SIR. The difference between each pair of models for all the shown results of SDR and SAR is statistically significant with $P$ values as follows. For SDR: $P(\mbox{DNN},\mbox{FCNN})=2\times10^{-5}$, $P(\mbox{DNN},\mbox{MR-FCNN})=2\times10^{-7}$, $P(\mbox{FCNN},\mbox{MR-FCNN})=0.0025$. For SAR: $P(\mbox{DNN},\mbox{FCNN})=7\times10^{-6}$, $P(\mbox{DNN},\mbox{MR-FCNN})=9\times10^{-7}$, $P(\mbox{FCNN},\mbox{MR-FCNN})=3\times10^{-5}$. We consider the difference between a pair of models statistically significant if $p < 0.05$, Wilcoxon signed-rank test \cite{Wilcoxon:45:icrm} and Bonferroni corrected \cite{hochberg:87:mcp}.    

\section{Conclusions}
In this work we proposed a new approach for monaural audio source separation (MASS). The new approach is based on using deep multi-resolution fully convolutional neural networks (MR-FCNN). The MR-FCNN learns unique multi-resolution patterns for each source and uses this information to separate the related components of each source from the mixed signal. The experimental results indicate that using MR-FCNN for MASS is a promising approach and can achieve better results than the feedforward neural networks and the single resolution FCNN. 

In our future work, we will investigate the possibility of applying the MR-FCNN on raw audio data (time domain signals) to extract multi-resolution time-frequency features that can represent the input data better than the STFT features. Some audio sources require higher resolution in the time than in the frequency, and other audio sources require the opposite resolution of that. By applying MR-FCNN on the raw audio data, we hope to extract useful features for each source according to its preferred time-frequency resolution which can improve the performance of any audio processing approach.  

\section{Acknowledgements}
 This work is supported by grants EP/L027119/1 and EP/L027119/2 from the UK Engineering and Physical Sciences Research Council (EPSRC).

% Below is an example of how to insert images. Delete the ``\vspace'' line,
% uncomment the preceding line ``\centerline...'' and replace ``imageX.ps''
% with a suitable PostScript file name.
% ------------------------------------------------------------------------
\vfill\pagebreak

%\section{REFERENCES}
%\label{sec:refs}

\bibliographystyle{IEEEbib}
\bibliography{strings,refs}

\begin{thebibliography}{10}

\bibitem{xiao:16:delmss}
X.~Zhang and D.~Wang,
\newblock ``{Deep} ensemble learning for monaural speech separation,''
\newblock {\em IEEE/ACM Trans. on audio, speech, and language processing}, vol.
  24, no. 5, pp. 967--977, 2016.

\bibitem{grais:14:ssrnmfmmse}
E.~M. Grais and H.~Erdogan,
\newblock ``{Source} separation using regularized {NMF} with {MMSE} estimates
  under {GMM} priors with online learning for the uncertainties,''
\newblock {\em Digital Signal Processing}, vol. 29, pp. 20--34, 2014.

\bibitem{Virtanen:07:msssbnmfwtcasc}
T.~Virtanen,
\newblock ``{Monaural} sound source separation by non-negative matrix
  factorization with temporal continuity and sparseness criteria,''
\newblock {\em IEEE Trans. on Audio, Speech, and Language Processing}, vol. 15,
  pp. 1066--1074, Mar. 2007.

\bibitem{Mimilakis:17:redasfcmsvs}
S.~I. Mimilakis, K.~Drossos, T.~Virtanen, and G.~Schuller,
\newblock ``{A recurrent} encoder-decoder approach with skip-filtering
  connections for monaural singing voice separation,''
\newblock in {\em arXiv:1709.00611}, 2017.

\bibitem{grais:17:tsscass}
E.~M. Grais, G.~Roma, A.~J. Simpson, and M.~D. Plumbly,
\newblock ``{Two} stage single channel audio source separation using deep
  neural networks,''
\newblock {\em IEEE/ACM Trans. on Audio, Speech, and Language Processing}, vol.
  25, no. 9, pp. 1469--1479, 2017.

\bibitem{Yuxuan:14:aspttsss}
Y.~Wang and D.~Wang,
\newblock ``{A structure-preserving} training target for supervised speech
  separation,''
\newblock in {\em Proc. ICASSP}, 2014, pp. 6148--6152.

\bibitem{Grais:17:descassdnn}
E.~M. Grais, G.~Roma, A.~J.R. Simpson, and M.~D. Plumbley,
\newblock ``{Discriminative} enhancement for single channel audio source
  separation using deep neural networks,''
\newblock in {\em Proc. LVA/ICA}, 2017, pp. 236--246.

\bibitem{chandna:17:massudcnn}
P.~Chandna, M.~Miron, J.~Janer, and E.~Gomez,
\newblock ``{Monoaural} audio source separation using deep convolutional neural
  networks,''
\newblock in {\em Proc. LVA/ICA}, 2017, pp. 258--266.

\bibitem{Venkataramani:17:nnacamss}
S.~Venkataramani, Y.~C. Subakan, and P.~Smaragdis,
\newblock ``{Neural} network alternatives to convolutive audio models for
  source separation,''
\newblock in {\em Proc. MLSP}, 2017.

\bibitem{Venkataramani:17:eessafe}
S.~Venkataramani and P.~Smaragdis,
\newblock ``{End-to-end} source separation with adaptive front-ends,''
\newblock in {\em Proc. WASPAA}, 2017.

\bibitem{grais:17:scasscda}
E.~M. Grais and Mark~D. Plumbly,
\newblock ``{Single} channel audio source separation using convolutional
  denoising autoencoders,''
\newblock in {\em Proc. GlobalSIP}, 2017.

\bibitem{Miron:17:msisscmcnn}
M.~Miron, J.~Janer, and E.~Gomez,
\newblock ``{Monaural} score-informed source separation for classical music
  using convolutional neural networks,''
\newblock in {\em Proc. ISMIR}, 2017.

\bibitem{Lim:17:hpsscae}
W.~Lim and T.~Lee,
\newblock ``{Harmonic} and percussive source separation using a convolutional
  auto encoder,''
\newblock in {\em Proc. EUSIPCO}, 2017.

\bibitem{fu:17:eewuedemofcnn}
S.~Fu, Y.~Tsao, X.~Lu, and H.~Kawais,
\newblock ``{End}-to-end waveform utterance enhancement for direct evaluation
  metrics optimization by fully convolutional neural networks,''
\newblock in {\em arXiv:1709.03658}, 2017.

\bibitem{Wenjie:16:uerfdcnn}
L.~Wenjie, L.~Yujia, U.~Raquel, and Z.~Richard,
\newblock ``{Understanding} the effective receptive field in deep convolutional
  neural networks,''
\newblock in {\em Proc. NIPS}, 2016, pp. 4898--4906.

\bibitem{Kawahara:16:mrtchpsltl}
J.~Kawahara and G.~Hamarneh,
\newblock ``{Multi-resolution-tract CNN} with hybrid pretrained and skin-lesion
  trained layers,''
\newblock in {\em Proc. MICCAI MLMI}, 2016, vol. 10019, pp. 164--171.

\bibitem{tang:12:mdbn}
Y.~Tang and A.~Mohamed,
\newblock ``{Multiresolution} deep belief networks,''
\newblock in {\em Proc. AISTATS}, 2012.

\bibitem{Zhang:17:HeartID}
Q.~Zhang, D.~Zhou, and X.~Zeng,
\newblock ``{HeartID:} a multiresolution convolutional neural network for
  {ECG}-based biometric human identification in smart health applications,''
\newblock {\em IEEE Access, Special Section on Body Area Networks}, pp.
  11805--11816, 2017.

\bibitem{Xue:17:emrtscar}
W.~Xue, H.~Zhao, and L.~Zhang,
\newblock ``{Encoding} multi-resolution two-stream cnns for action
  recognition,''
\newblock in {\em Proc. ICONIP}, 2016, pp. 564--571.

\bibitem{Naderi:17:mrcnnrsr}
N.~Naderi and B.~Nasersharif,
\newblock ``{Multiresolution} convolutional neural network for robust speech
  recognition,''
\newblock in {\em Proc. ICEE}, 2017.

\bibitem{se:16:fcnnse}
S.~R. Park and J.~W. Lee,
\newblock ``{A} fully convolutional neural network for speech enhancement,''
\newblock in {\em Proc. Interspeech}, 2017.

\bibitem{Mengyuan:16:mrcdasr}
M.~Zhao, D.~Wang, Z.~Zhang, and X.~Zhang,
\newblock ``{Music} removal by convolutional denoising autoencoder in speech
  recognition,''
\newblock in {\em In proc. APSIPA}, 2016.

\bibitem{Klapuri:07:spmmt}
M.~Davy A.~Klapuri,
\newblock {\em {Signal} Processing Methods for Music Transcription},
\newblock Springer, 2007.

\bibitem{emad:12:avsrwbmscss}
E.~M. Grais, I.~S. Topkaya, and H.~Erdogan,
\newblock ``{Audio-Visual} speech recognition with background music using
  single-channel source separation,''
\newblock in {\em Proc. SIU}, 2012.

\bibitem{emad:13:stpemenbscss}
E.~M. Grais and H.~Erdogan,
\newblock ``{Spectro-temporal} post-enhancement using {MMSE} estimation in
  {NMF} based single-channel source separation,''
\newblock in {\em Proc. InterSpeech}, 2013.

\bibitem{ono:15:tsisec}
N.~Ono, Z.~Rafii, D.~Kitamura, N.~Ito, and A.~Liutkus,
\newblock ``{The 2015 signal separation evaluation campaign},''
\newblock in {\em Proc. LVA/ICA}, 2015, pp. 387--395.

\bibitem{vincent:06:pmi}
E.~Vincent, R.~Gribonval, and C.~Fevotte,
\newblock ``{Performance} measurement in blind audio source separation,''
\newblock {\em IEEE Trans. on Audio, Speech, and Language Processing}, vol. 14,
  no. 4, pp. 1462--69, July 2006.

\bibitem{Emad:16:scassdnne}
E.~M. Grais, G.~Roma, A.~J.~R. Simpson, and M.~D. Plumbley,
\newblock ``{Single} channel audio source separation using deep neural network
  ensembles,''
\newblock in {\em Proc. 140th Audio Engineering Society Convention}, 2016.

\bibitem{grais:16:cmescassdnn}
E.~M. Grais, G.~Roma, A.~J.~R. Simpson, and M.~D Plumbley,
\newblock ``{Combining} mask estimates for single channel audio source
  separation using deep neural networks,''
\newblock in {\em Prec. InterSpeech}, 2016.

\bibitem{adam:14:amso}
D.~P. Kingma and J.~Ba,
\newblock ``{Adam A} method for stochastic optimization,''
\newblock in {\em Proc. ICLR}, 2015.

\bibitem{chollet2015keras}
F.~Chollet,
\newblock ``Keras, https://github.com/fchollet/keras,'' 2015.

\bibitem{Wilcoxon:45:icrm}
F.~Wilcoxon,
\newblock ``{Individual} comparisons by ranking methods,''
\newblock {\em Biometrics Bulletin}, vol. 1, no. 6, pp. 80--83, 1945.

\bibitem{hochberg:87:mcp}
Y.~Hochberg and A.~C. Tamhane,
\newblock {\em {Multiple} Comparison Procedures},
\newblock John Wiley and Sons, 1987.

\end{thebibliography}

\end{document}